\documentclass{emulateapj}
\usepackage{multirow}
\usepackage{color}
\usepackage{amsmath,amssymb}

\begin{document}
\submitted{}

\title{On the influence of the Illuminati in astronomical adaptive optics}
\date{1 April 2012}
\shorttitle{AO Illuminati}

\author{Katie M. Morzinski$^{1,*}$ and Jared R. Males$^{1,\ddagger}$}
\address{$^1$Steward Observatory, University of Arizona, 933 N. Cherry Ave., Tucson, AZ 85721, USA}
\email{$^*$Fellow.  Contact ktmorz@arizona.edu for general queries.}
\email{$^\ddagger$Fellow Fellow.  Contact jrmales@as.arizona.edu if discretion required.}

\begin{abstract}
Astronomical adaptive optics (AO) has come into its own.  Major O/IR telescopes are achieving diffraction-limited imaging; major facilities are being built with AO as an integral part.
To the layperson, it may seem that AO has developed along a serpentine path.
However, with a little illumination, the mark of Galileo's heirs becomes apparent in explaining the success of AO.
\end{abstract}
\maketitle

\section{Introduction}
It is clear that the Illuminati are alive and well in modern times \citep{angelsanddemons}.
For instance, it is well known that pop stars Britney Spears and Lady Gaga have been aided in their astronomical rise to the top by the Illuminati\footnote{See the comments for any video at youtube.com/britneyspears and youtube.com/ladygaga}
\citep{youtube}.
The secret to success in ground-based diffraction-limited astronomical imaging is less well known.
Toward that end, we have conducted an exhaustive study into the history and nature of the clandestine organization directing all development in adaptive optics: the AO Illuminati.

\subsection{Background}
The power of adaptive optics (AO) was demonstrated yet again in recent weeks, in a paper splash reporting the first science results \citep{2012arXiv1203.2638C,2012arXiv1203.2735E,2012arXiv1203.2619R,2012arXiv1203.2615S} from the LBT's first-light AO system \citep{2011SPIE.8149E...1E}.
But how did astronomical AO develop, and what delayed astronomers from achieving 100 nm rms wavefront error on-sky?

AO as conceived by \cite{babcock1953} was realized by the U.S.\ military in the 1970s \citep{tyson1994,tyson1997}.
Astronomers prototyped the first AO systems in a non-military setting in the late 1980s \citep{babcock1990,rousset1990}, in bittersweet triumph at having independently discovered the secrets of AO:
\begin{quote}
By contrast, the US military is thought to have been working for years on adaptive optics technology for the purpose of satellite tracking and laser weaponry \emph{without sharing any of its knowledge with astronomers}.
\citep{dickman1989} [Emphasis added.]
\end{quote}
Sensing trouble afoot, the AO Illuminati chose to de-classify a select stream of research in 1991 \citep{collins1992}, in order to guard their secrets by controlling their release.
Indeed, astronomy is funding-limited, and the AO Illuminati could still exercise control over AO development through financial avenues:
\begin{quote}
The most ambitious civilian programme was dropped recently by the National Optical Astronomy Observatories in Tucson, Arizona, because of \emph{lack of money and personnel}.
\citep{dickman1989} [Emphasis added.]
\end{quote}
John W.\ Hardy, author of the AO Bible \citep{hardy}, also bemoaned slow progress in astronomical AO:
\begin{quote}
Adaptive optics for image compensation in astronomy has not fulfilled its early promise.  When the first experiments were made about 15 years ago [\textit{e.g.} \citep{bridges1974,buffington1977observatory,hardy1977}], showing that real-time compensation of turbulence-degraded images was possible using the light from the object itself as the reference, we had a vision of a golden age for ground-based astronomy in which diffraction-limited images would come pouring out of ever-larger telescopes.  \emph{This has not happened.}   Astronomers are resourceful people and \emph{it is certainly not due to any lack of incentive on their part}.
\citep{hardy1989} [Emphasis added.]
\end{quote}

What prevented the ``Golden Age'' from immediately occurring?
The hand of the Illuminati is clear, but what were their goals?  Did they for some reason seek to limit the progress of astronomical AO, or were they attempting to ensure that the Golden Age would happen when the time was right?
Today, astronomers of all walks must trade-off discovery space, resource sharing, and risk reduction in pursuit of funding \citep{astro2010}.  It may be, however, that AO scientists must deal with an even harsher discriminant: the secret hand of the Illuminati guiding the development of AO.  In what follows we will explore the nature of this guidance.

\subsection{Motivation}
The present study was motivated by an offhand remark to KMM by [redacted], lamenting the fact that the [redacted] project could not get funding due to having selected a Curvature system for wavefront sensing.  The stated reason for this problem was that ``the AO Illuminati do not like Curvature sensors.'' \footnote{True story.}

This argument is not without merit --- as one of the main alternatives to Curvature, the Pyramid wavefront sensor (WFS) is the first to produce a Dark Hole on an 8 meter telescope on-sky \citep{esposito2010}.  (The ``Prophecy of the Dark Hole'' is discussed in detail below.)
Of note is the interesting correspondence between a Pyramid WFS and the rotating Knife-Edge sensor of Babcock's original conception (Babcock 1953, Fig.~1).
Indeed, it is common for students of the Pyramid who do not actually understand how it works to state with feigned confidence ``it's just the knife-edge test'' (these same students can seldom coherently explain how the knife-edge works either).

Thus we see that AO has come full-circle from the Knife-Edge to the Pyramid.  Will the frustrations of generations of astronomers be put to rest as we finally enter AO's Golden Age?
To understand the answer, we must understand the agency controlling AO development.
It is therefore the goal of this work to explore the existence and essence of the AO Illuminati.
 
\section{Symbology of the AO Illuminati}
It is well known that the main cohort of the Illuminati traffic in mysterious symbols and signs \citep{angelsanddemons,youtube}.  The AO Illuminati are no different.  Here we reveal, for the first time in print, the symbology of the AO Illuminati.

\subsection{The Reversed Deformable Mirror}
\begin{figure}
\begin{center}
\includegraphics[width=0.9\linewidth]{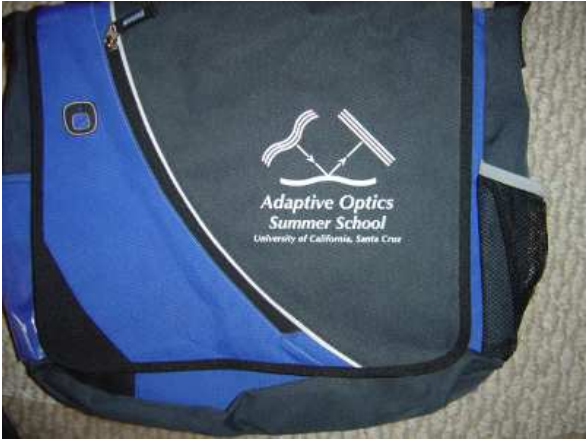}
\end{center}
\caption{The book bag given out during the AO Summer School at CfAO in recent years.  The deformable mirror is printed such that it would amplify rather than correct the incoming wavefront.  We believe that the so-called ``Reversed Deformable Mirror'' may be the first known symbol of the AO Illuminati. \\
\label{fig:symbol_bag}}
\end{figure}

At the 2009--2010 CfAO AO Summer Schools, attendees received a book bag with the logo of the CfAO on it.  The logo was printed with what seems like a subtle error.  The deformable mirror was printed backwards, such that it would function as an amplifier of the incoming wavefront rather than a corrector.  A photo of one specimen is shown in Fig.~\ref{fig:symbol_bag}.  As one can see (if one wasted enough time with it) there does not appear to be a simple explanation such as a mistaken reflection in Photoshop, and the commonly-touted ``Non-Common Path'' defense is unsatisfactory as well.  Given all the other evidence we have been unable to gather, we are forced to conclude that the Reversed Deformable Mirror may be the first known symbol of the AO Illuminati.

\subsection{The Prophecy of the Dark Hole}
\begin{figure}
\begin{center}
\includegraphics[width=0.7\linewidth]{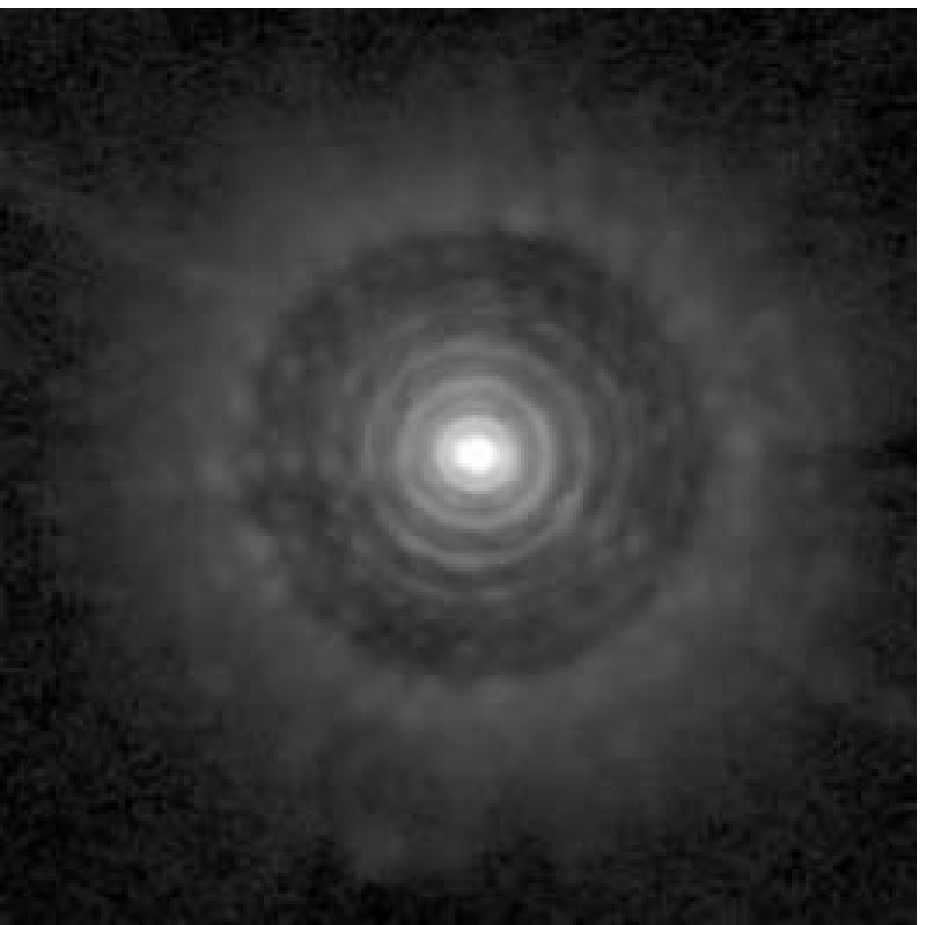}
\end{center}
\caption{Dark Hole, LBT, \cite{esposito2010}.
\label{fig:darkhole}}
\end{figure}

It is rumored that in some copies of Hardy (including the ones that went missing from the CfAO library around the time that KMM entered grad school there) the following words can be found in the margin of the first page of chapter 5: ``they shall be known by their Dark Hole."  We can only surmise that this purported prophecy indicates that members of the AO Illuminati will make themselves known by producing the mythical Dark Hole in the PSF of their instruments.  (See Fig.~\ref{fig:darkhole} for the Dark Hole produced by the LBT AO system that caused such a stir at SPIE 2010 in San Diego.)
Furthermore, since Chapter 5 of \cite{hardy} concerns ``Optical Wavefront Sensors,'' this seems to be a sign that the key to the Illuminati's stranglehold on Astronomical AO is the wavefront sensor.

\subsection{The Moustache}
Under debate.  See Fig.~\ref{fig:claire}.

\section{Conventions of the Illuminati}

\subsection{Fiducial wavelength}
Some AO astronomers quote $r_0$ at 500 nm, others at 550 nm, and others refuse to commit to a wavelength.
It may be that the cop-out of simply stating ``V band'' is in fact an Illuminati tactic to sow confusion within the AO community.

\subsection{Performance metric}
Strehl is often used as an AO performance metric by the uninitiated.
However, true Illuminati-trained astronomers prefer wavefront error (WFE),
as any fool with a long-wavelength filter can obtain a high Strehl.

\subsection{Pronunciation}
Americans and Canadians pronounce Strehl as either ``Strell'' or ``Strayl.''  The former pronunciation is found within the US AO community in Arizona and California; the latter in Hawaii though there may be a Chicago-Roddier connection (Mark Chun, personal communication, 2012).  Linguists are studying the history of this word, in order to trace the pathway taken by the Illuminati as they took hold of the global AO community.

\section{Members of the Illuminati}
Here we document our attempt to determine which Astronomers are or
may be members of the Illuminati.  Due to funding constraints, we have
so far been forced to concentrate mainly on people associated with
CfAO or CAAO.  Future work, once fully funded, will explore other Illuminati
strongholds such as Hawaii, Caltech, HIA, Durham, ESO, etc.

\subsection{Known Members}

\subsubsection{Claire E.\ Max}
\begin{figure}
\begin{center}
\includegraphics[width=3.4cm]{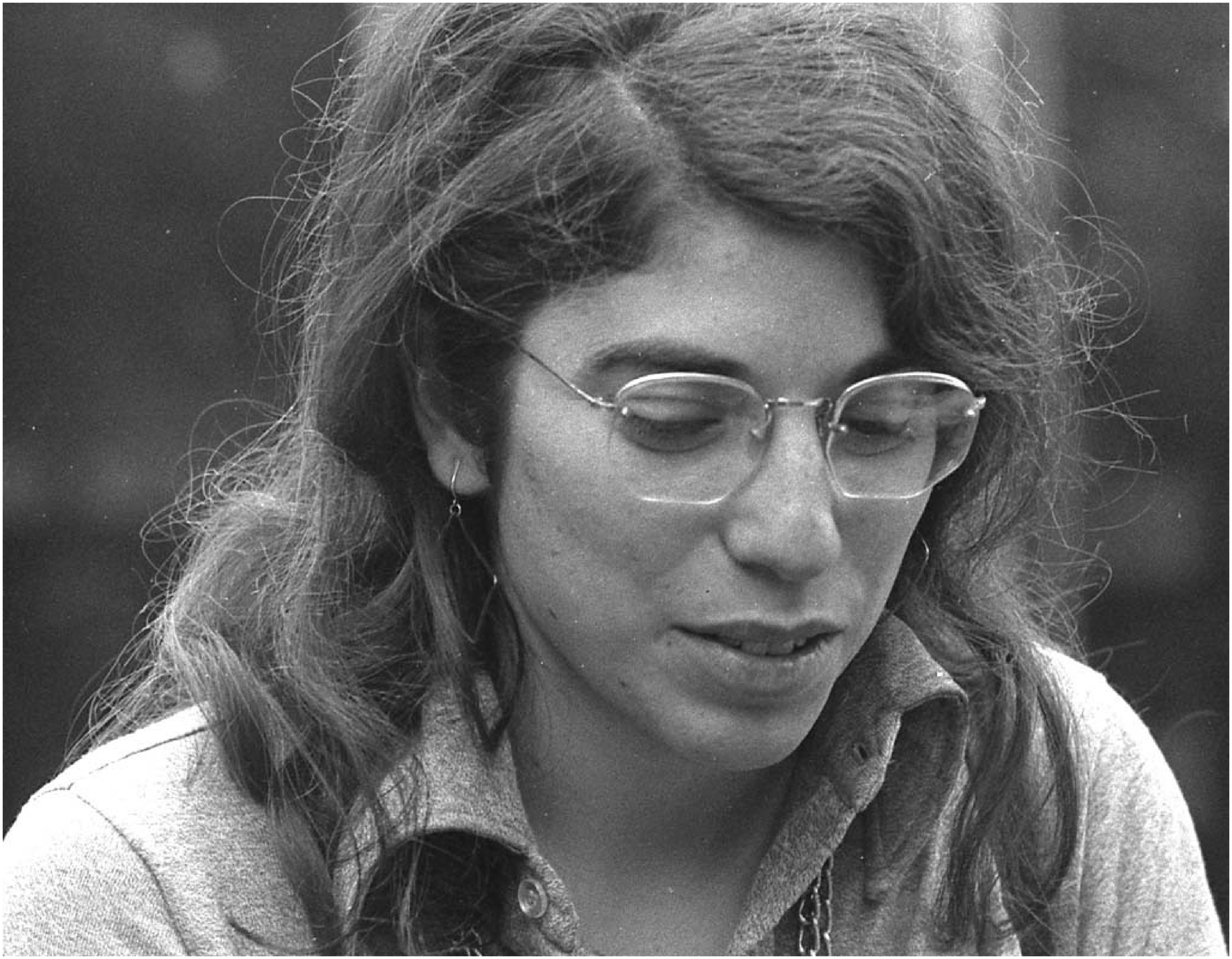}
\includegraphics[width=3.6cm]{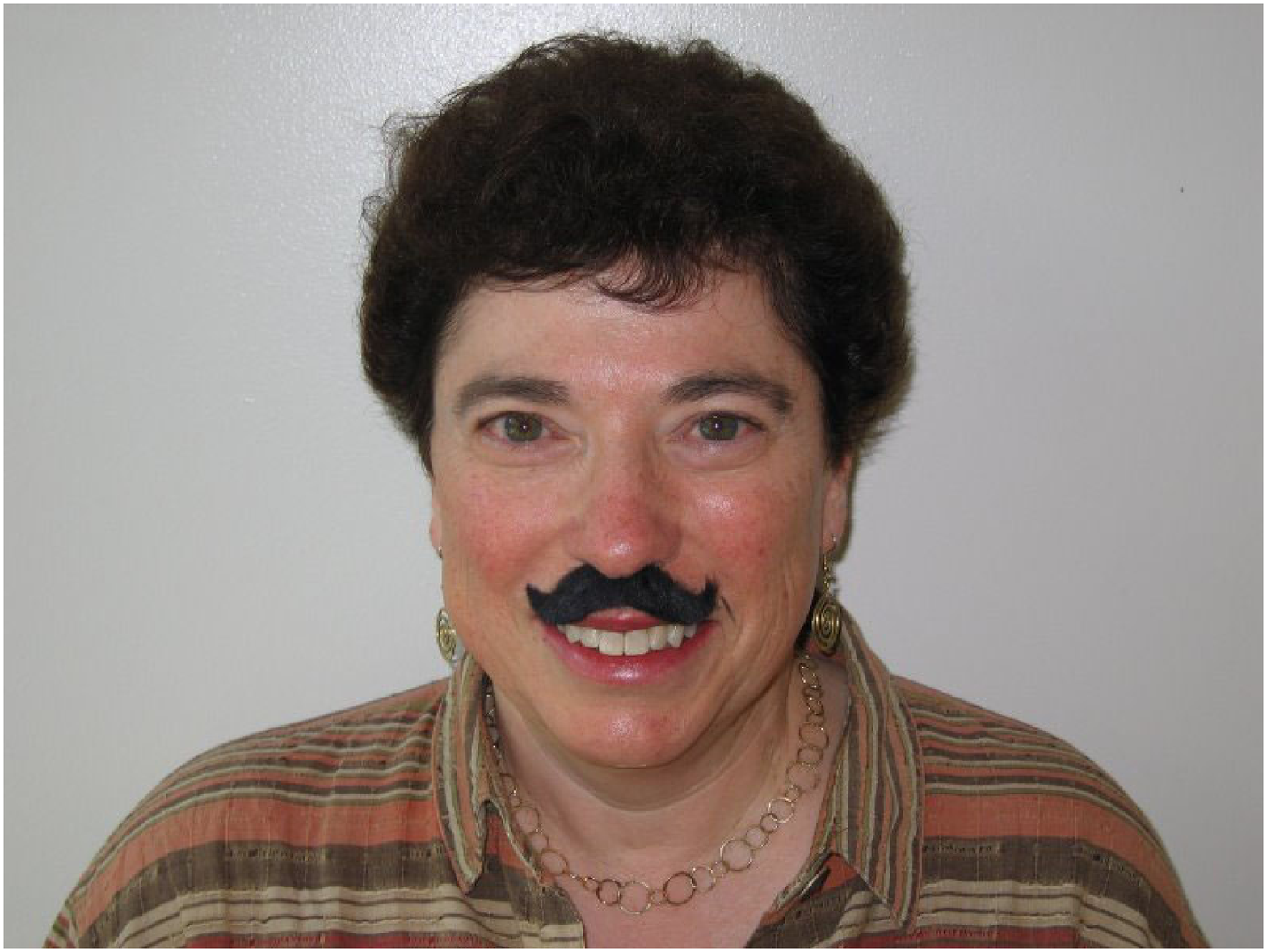}
\end{center}
\caption{Claire Max, before (Left) and after (Right) becoming a fixture of the AO Illuminati.  As Max is neither a hipster nor a melodrama villain, we conclude the moustache to be an inexplicable Illuminati mark.
\label{fig:claire}}
\end{figure}

Director of the Center for Adaptive Optics at the University of California.
Max was the very person cited in reference to the AO Illuminati preventing the [redacted] project from proceeding with Curvature AO.
This unguarded admission forces us to conclude that Claire Max is in fact a member of the AO Illuminati, though she is currently one of only two people we are sure of.  She is shown in Fig.~\ref{fig:claire} looking highly suspicious indeed.

As if that weren't enough, we note the following: the French word \textit{clair} means bright, light, or clear.  Therefore, \textit{Clair Max} roughly translates to maximum brightness or maximum light.  Oh, and speaking of lasers, Claire Max co-invented the sodium laser guide star --- Coincidence?

\subsubsection{Simone Esposito}
\begin{figure}
\begin{center}
\includegraphics[width=0.7\linewidth]{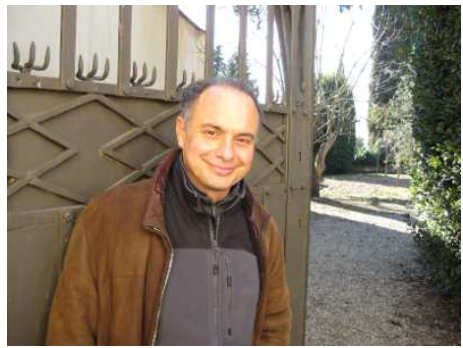}
\end{center}
\caption{Simone Esposito apparently about to lock the gate to Galileo's house, Florence, Feb.\ 2012.  He would not continue until we left the premises.  See text for the full implications to this photograph.
\label{fig:simone}}
\end{figure}

Director of the AO group at Arcetri Observatory in Florence.
Achieved 100 nm rms WFE on sky \citep{esposito2010}.
Known as ``the Lord of the Rings'' in certain sections of Steward Observatory, a reference to the Airy rings produced by LBTAO (see Fig.~\ref{fig:darkhole}).
His Dark Hole is round, like the eye of Sauron.
Evidence shows that Esposito can access a key to Galileo's house (see Fig \ref{fig:simone}).  We know from \cite{angelsanddemons} that Galileo was one of the original members of the Illuminati.  We'll say it again: the man has a key to Galileo's frickin' house. QED.

\subsection{Suspected Members}
\subsubsection{Norbert Hubin}
Director of AO at ESO.
Has attended the CfAO Fall Retreat and acted very convivially.  A collaborator of Claire Max's --- or a spy?
We have barely begun to plumb the depths of the European AO community, and it is likely that an entirely parallel Illuminati exists there, perhaps with Hubin as its head.

\subsubsection{Roger Angel and/or Jerry Nelson}
Former Director of CAAO at the University of Arizona and/or the CfAO at the University of California.
Shared 2010 Kavli prize for astronomy for ``contributions to the development of giant telescopes.''
Build humongous mirrors like nobody's business.
Clear suspects.

\subsection{Persons To Watch}

\begin{figure}
\begin{minipage}[b]{0.44\linewidth}
\centering
\includegraphics[width=3.9cm]{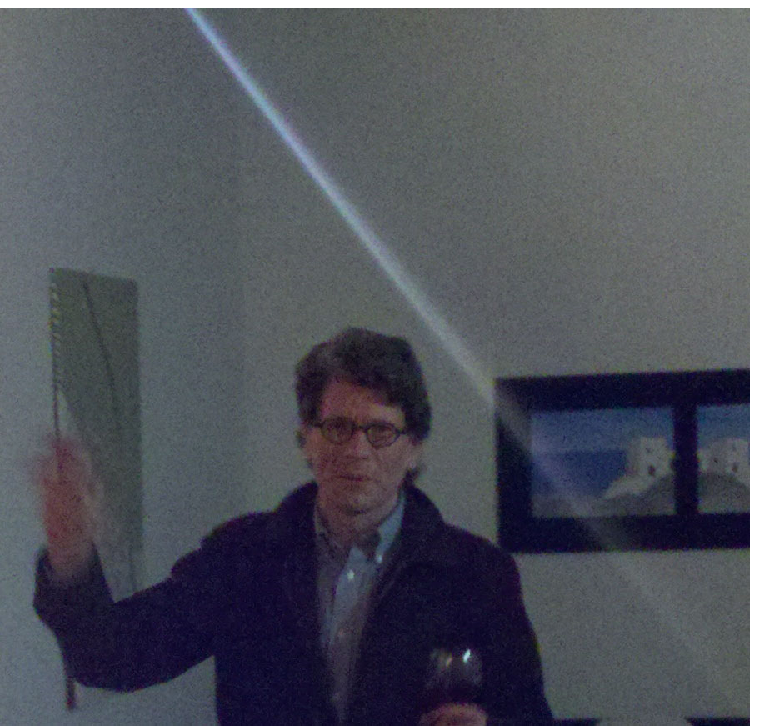}
\caption{Laird Close marked by a beam of light.
\label{fig:laird}}
\end{minipage}
\hspace{0.5cm}
\begin{minipage}[b]{0.44\linewidth}
\centering
\includegraphics[width=2.6cm]{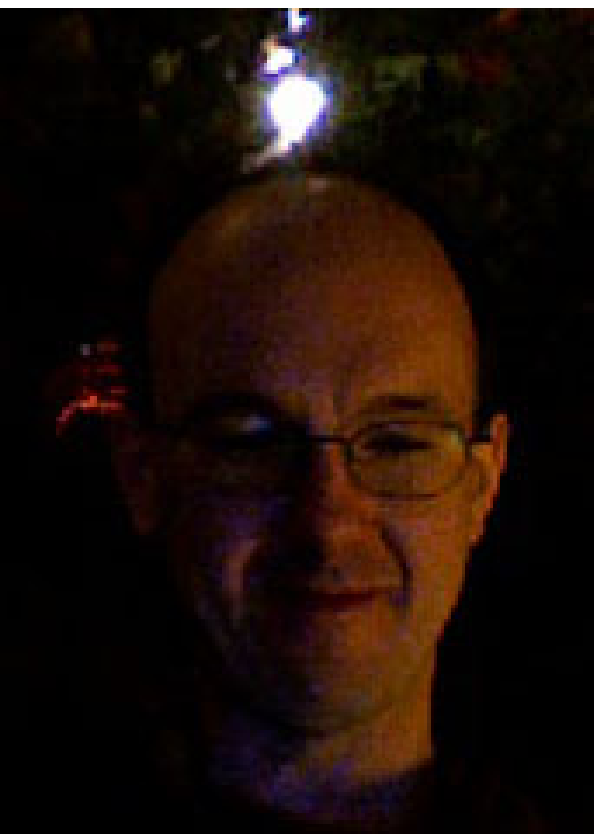}
\caption{Bruce Macintosh marked by a beacon of light.
\label{fig:bmac}}
\end{minipage}
\end{figure}

\subsubsection{Laird M.\ Close}
Laird Close is among the few astronomers to have used all three of Curvature, Shack-Hartmann, and Pyramid wavefront sensors on-sky.  He has reported to one of us (JRM) that he was present on the first pilgrimage by Arizona astronomers to the Starfire Optical Range in New Mexico (USA) to be indoctrinated into the AO world.  Perhaps the most interesting note on his record is the apparent trajectory of the Magellan AO system.  Generally ignored by the AO community, MagAO continues to soldier on.\footnote{Mostly they make videos and post to their blog at http://visao.as.arizona.edu/.}  The success or failure of this project may tell us Close's status within the Illuminati in the near future.
Shown with suspicious illumination in Fig.~\ref{fig:laird}.

\subsubsection{Bruce A.\ Macintosh}
The director of CfAO's ExAO Theme and the PI of GPI, Macintosh co-discovered the Illuminati-anointed planets HR 8799 bcde \citep{hr8799,marois2010}.
Although Macintosh appears to reside in the spatial domain, his mind is often in the Fourier domain ---
he determined how to create the Dark Hole with a Shack-Hartmann, using the Spatially-Filtered WFS \citep{poyneer_macintosh}.
Once GPI produces its first on-sky Dark Hole, we will know for sure.
He is shown with suspicious illumination in Fig.~\ref{fig:bmac}.

Macintosh often states that his most important contribution to Keck AO was hiring Marcos van Dam.
Van Dam has since returned to New Zealand, started a company called Flat Wavefronts (but the question is, How flat RMS?), and is a suspect for expanding the Illuminati into Oceania.

\subsubsection{Phil Hinz}
Hinz is Angel's anointed successor at the University of Arizona.  His meteoric rise to director of CAAO is nothing short of remarkable, and he is PI of a major AO effort (the LBTI).  He is also inextricably linked to MMTAO, and is in fact generally credited by Laird Close with first proposing that the LBTAO ASM would work well on the Magellan 6.5 meter apertures.  Watch this man.

\subsubsection{Olivier Guyon}
Due to his involvement with Curvature AO
(the dislike of Curvature by the Illuminati forms the entire basis of this paper) this is an interesting case.
It has also been observed that Guyon seldom remains in the same time zone for longer than $\sim$24 hours (judging by the pile of expended boarding passes he keeps on his desk at SO, this should be considered an upper limit), a possible countermeasure on his part.  If he is not Illluminati, he may instead be one of its targets.  Watch out O.\ G.

\section{Controls}
What if the authors themselves are AO Illuminati or agents thereof? Given our close involvement with several of the people and projects named above we feel that neither of us should be trusted without a great deal of thought.  Below we provide a guide to mistrusting our results, in the form of several hypotheses that one should consider.

\subsection{Hypothesis 1: JRM is Illuminati; KMM is not}
Given his lowly status as a graduate student at Steward Observatory (see \cite{steward_lpl} for how lowly) and his aversion to carrying the symbol of the Reversed Deformable Mirror, this seems the least likely case.  However, given his association with certain sections of the US national security apparatus in the not so distant past, one should not rule this out entirely.

\subsection{Hypothesis 2: KMM is Illuminati; JRM is not}
We judge this to also be unlikely.  Though she has achieved a slightly more exalted status than JRM, it is doubtful she would have been admitted at such a young age, even if she is the only known astronomer to have used all three of piezo, MEMS, and ASM wavefront correctors on-sky.

\subsection{Hypothesis 3: Neither JRM nor KMM are Illuminati}

Probably the most likely state of the authors.  If you, Dear Reader, agree, then we note here that you are hereby forced to acknowledge the breathtaking level of intellectual courage the authors are displaying in publishing our work on this topic.

\subsection{Hypothesis 4: Both JRM and KMM are Illuminati}

This terrifying prospect comes in two flavors.

\subsubsection{Both are, but at least one of us does not know about the other}

In this case, a deeper Illuminati plot is afoot.  It would have to be so important that our Illuminati betters could not admit the truth to either of us but are still allowing this publication.  This is a somewhat unsettling possibility.  What could it be?

\subsubsection{Both are, and we know it}

A.K.A. the ``this entire paper is bullshit'' hypothesis.  Frankly, you'll have to decide for yourself.

\section*{Acknowledgments}

We would like to thank the anonymous referee, Andrew Joseph Imonode Skemer (askemer@as.arizona.edu), though he rejected this paper out-right and has since refused to review further submissions.

You know who you are.

\bibliography{20120401.bbl}
\end{document}